\documentclass[twocolumn,showpacs,preprintnumbers,amsmath,amssymb]{revtex4}
\usepackage{graphicx} % Include figure files
\usepackage{dcolumn} % Align table columns on decimal point
\usepackage{bm} % bold math

\begin{document}

%\preprint{APS/xxx-xxx}

\title{Spin Polarized Current in the Ground State of \\ 
       Superconductor - Ferromagnet - Insulator Trilayers.}

\author{M. Krawiec}
 \email{m.a.krawiec@bristol.ac.uk}
\author{B. L. Gy\"orffy}
\author{J. F. Annett}
\affiliation{H. H. Wills Physics Laboratory, 
 University of Bristol, Tyndal Ave., Bristol BS8 1TL, UK}

\date{\today}

\begin{abstract}
We study the ground state properties of a superconductor - ferromagnet -
insulator trilayer on the basis of a Hubbard Model featuring exchange splitting
in the ferromagnet and electron - electron attraction in the superconductor. We 
solve the spin - polarized Hartree - Fock - Gorkov equations together with the 
Maxwell's equation (Ampere's law) fully self-consistently. For certain values 
of the exchange splitting we find that a spontaneous spin polarized current is 
generated in the ground state and is intimately related to Andreev bound 
states at the Fermi level. Moreover, the polarization of the current strongly 
depends on the band filling.
\end{abstract}
\pacs{72.25.-b, 74.50.+r, 75.75.+a}
%\keywords{proximity effect}

\maketitle

%%%%%%%%%%%%%%%%%%%%%%%%%%%%%%%%%%%%%%%%%%%%%%%%%%%%%%%%%%%%%%%%%%%%%%%%%%%%%%

{\center
\section{\label{sec1} INTRODUCTION}
}
Recently, the proximity effect between a superconductor ($SC$) and a 
ferromagnet ($FM$) has attracted much attention because, due to advances in 
materials growth and fabrication techniques \cite{Tedrow}, well controlled 
structures, in which it can occur became available. Such $FM$/$SC$ hybrid 
structures are important from the point of view of their intrinsic scientific 
interest, as they allow the study of the interplay between ferromagnetism and 
superconductivity \cite{Berk} as well as of device applications in such areas 
of technology as magnetoelectronics \cite{Bauer} or quantum computing 
\cite{Blatter}.

In the present context the proximity, on one hand, means the leakage of 
superconductivity into the ferromagnet, and on the other, the spin 
polarization of the superconductor. In the normal metal - superconductor 
systems the proximity effect has been studied over 30 years, and is now well 
understood \cite{LambertRaimondi} to be governed by the Andreev reflection 
\cite{Andreev} processes, in which an electron at the interface is reflected 
as a hole while a Cooper pair is transfered into superconductor. This mechanism 
makes the movement of electrons and holes in the normal metal highly coherent. 

If the normal metal is replaced by a ferromagnet a number of new phenomena may
occur. For instance one may expect a version of the $FFLO$ state, predicted in 
sixties for a bulk superconductor in a strong spin exchange field, by Fulde and 
Ferrell \cite{FuldeFerrell} and Larkin and Ovchinikov \cite{LarkinOvchinnikov}, 
to be realized. In the bulk, as it is well known, the exchange field tends to 
polarize the conduction electrons and holes. In particular if these electrons 
are Cooper paired, naively one would expect that either the exchange field is 
too week to break the pairs, or it leads to the first order phase transition to 
the normal state. However, as was demonstrated in Ref. 
\cite{FuldeFerrell,LarkinOvchinnikov}, for certain values of the exchange field 
a new superconducting depairing state is realized through first order phase 
transition from the $BCS$ state and it transforms continuously, by a second 
order phase transition, into the normal state as the strength of the exchange 
field is increased. This $FFLO$ state has a spatially dependent order parameter 
corresponding to the nonzero center of mass motion of the Cooper pairs. Another 
novel feature of this state is a current flowing in the ground state. The 
unpaired electrons tend to congregate at one portion of the Fermi surface so 
a quasiparticle current is produced. In order to satisfy the Bloch theorem: no 
current in the ground state, a supercurrent, generated by the nonzero value of 
the pairing momentum, flows in opposite direction, and the total current is 
zero. 

Interestingly, similar oscillations of the pairing amplitude (Cooper pair 
density) have been predicted \cite{Buzdin82}-\cite{Demler} in a ferromagnet 
proximity coupled to a superconductor. It turns out that these oscillations 
are responsible for the oscillatory dependence of the critical temperature 
on the thickness of the ferromagnetic slab \cite{Buzdin90,Radovic}.
This effect as well as the corresponding oscillations of the density of states 
at the Fermi level \cite{Prokic,Vecino}, have been observed experimentally 
\cite{Wong,Kontos}. In this paper we investigate further the ramifications of 
these interesting phenomena.

One of these is the formation of so called $\pi$ junction state \cite{Buzdin91} 
which is also a fingerprint of the oscillatory behaviour of the pairing 
amplitude. The $\pi$-state effect has been extensively studied in connection 
with the high-$T_c$ superconductors, where due to spatially inhomogeneity of 
the order parameter, the Cooper pair wave function can change sign at the 
interface and this leads to the formation of a zero - energy mid - gap state. 
Remarkably, this zero - energy state is unstable to the occurrence of a 
spontaneous current flowing parallel to and within a coherence length of the 
interface (see eg. \cite{Kashiwaya} and references therein). In what follows 
we report on our finding similar spontaneous currents but of different origin 
in a $SC$/$FM$/$I$ trilayer.

We also find zero energy bound states due to the finite size of the $FM$ slab. 
In fact even for a $N$/$SC$ heterostructure, in agreement with 
\cite{deGennes-SJames}, we have found states within the $SC$ gap. Although the 
energy of these states depend on the thickness of the normal metal, they never 
reach zero, namely $\varepsilon_F$. Interestingly, when the normal metal is 
replaced by a ferromagnet, these bound states split, and for certain values of 
the exchange field, they cross the Fermi level \cite{Vecino}. As we shall show, 
this circumstance leads to a current flow. Clearly, the spontaneous current is 
strictly related to the zero energy states, as in the case of the high - $T_c$ 
superconductors, but the origin of these states is completely different. As 
mentioned above, there are two competing parts to the total current in the 
$FFLO$ state. One is related to the unpaired electron movement and the other is 
a supercurrent. So one may expect, that in the case of the $FM$/$SC$ 
heterostructure, these parts could be spatially unbalanced creating regions of 
net current flow. Indeed this is what we have discovered.

A brief report of our main results has already been published \cite{KGA}. Here 
we wish to present a more detailed and systematic study of the zero energy
Andreev bound states and the corresponding spontaneous current in the $FM$/$SC$ 
heterostructures. The paper is organized as follows: In Sec. \ref{sec2} the 
simple model which allows for self-consistent description of the $FM$/$SC$ 
heterostructure is introduced. We also derive equations which have to be solved 
self-consistently. In particular, these include the Maxwell's equation 
(Ampere's law) for the coupling of the current to the magnetic field. Then we 
describe some technical details concerning the principal layer technique, which 
allows us to describe a semi-infinite superconductor. In Sec. \ref{sec3} the 
nature of the Andreev bound states in the ferromagnet is discussed. The 
spontaneously generated current and corresponding magnetic field in the ground 
state are studied in the Sec. \ref{sec4}. In Sec. \ref{sec5} we provide some 
suggestions of how one might observe the spontaneous current and its 
polarization experimentally. Finally, we conclude in Sec. \ref{sec6}.

\vspace{1cm}
%%%%%%%%%%%%%%%%%%%%%%%%%%%%%%%%%%%%%%%%%%%%%%%%%%%%%%%%%%%%%%%%%%%%%%%%%%%%%%

{\center
\section{\label{sec2} THE MODEL}
}

{\center
\vspace{0.5cm}
\subsection{\label{subsec2.a} Negative $U$ Hubbard Model with exchange 
            splitting}
}

\begin{figure}[h]
 \resizebox{8cm}{!}{
  \includegraphics{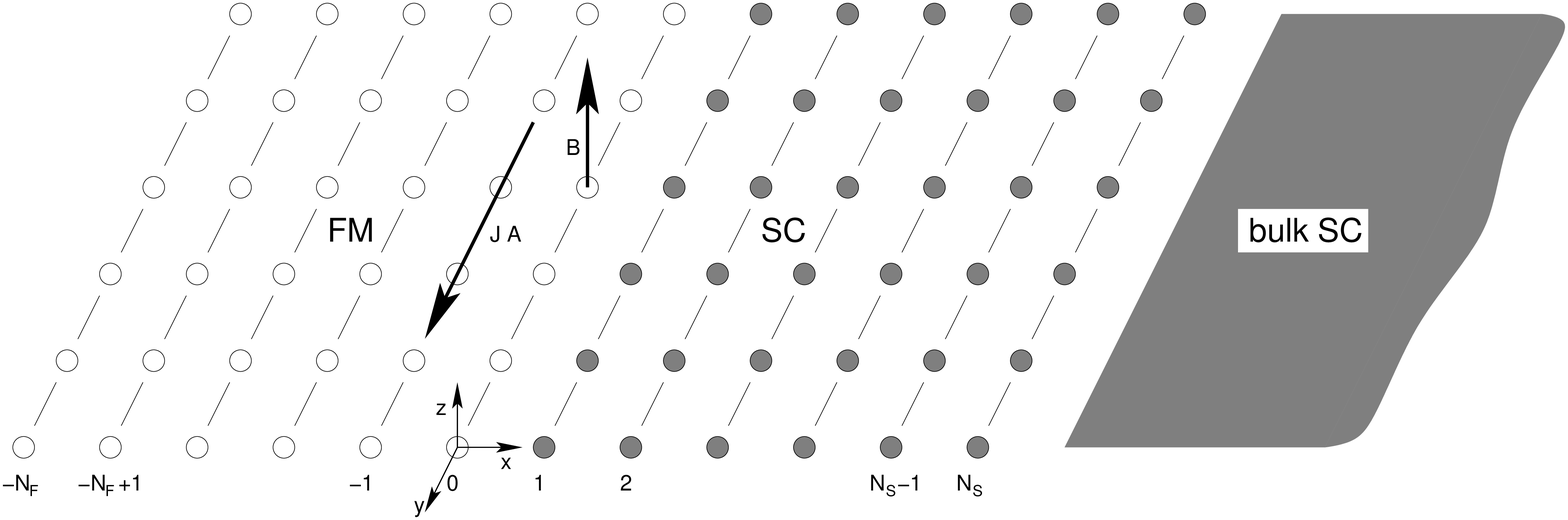}}
 \caption{\label{Fig1} Schematic view of the (finite thickness) ferromagnet - 
          semi-infinite superconductor heterostructure. Directions of the 
	  magnetic field ($B$) as well as vector  potential ($A$) and current 
	  ($J$) are indicated.}
\end{figure}

To study spontaneous currents and their polarization in the ferromagnet - 
superconductor heterostructure we have adopted a single orbital, nearest 
neighbour hopping Hubbard model with negative $U$ on the $SC$ side and zero 
$U$ otherwise. Additionally we have assumed the site energies
$\varepsilon_{i\sigma}$ to be exchange split in the ferromagnet and spin 
independent in the superconductor. Moreover, the simplest geometry allowing 
for a current flow is the $2D$ system, shown in the Fig.\ref{Fig1}, where 
the magnetic field in one direction, the vector potential and the current in 
another and the spatial modulation in a third orthogonal direction are 
explicitly indicated.

In short, our model Hamiltonian is given in the form:
\begin{eqnarray}
 H =  \sum_{ij\sigma} [t_{ij} + (\varepsilon_{i\sigma} - \mu) \delta_{ij}] 
      c^+_{i\sigma} c_{j\sigma} + 
      \sum_{i\sigma} \frac{U_i}{2} \hat n_{i\sigma} \hat n_{i-\sigma}
 \label{Hamiltonian}
\end{eqnarray}
where, in the presence of a vector potential $\vec{A}(\vec{r})$, the hopping
integral is given by $t_{ij} = - t e^{-i e \int_{\vec{r}_i}^{\vec{r}_{j}} 
\vec{A}(\vec{r}) \cdot d\vec{r}}$ for nearest neighbour lattice sites, whose 
positions are $\vec{r}_i$ and $\vec{r}_j$, and zero otherwise. The site 
energies $\varepsilon_{i\sigma}$ are $0$ on the superconducting side and equal 
to $\frac{1}{2} E _{ex}\sigma$ on the ferromagnetic side, $\mu$ is the chemical 
potential, the on-site interaction $U_i$ is $U_S < 0$ in the superconductor and 
zero elsewhere, $c^+_{i\sigma}$, ($c_{i\sigma}$) are the usual electron 
creation (annihilation) operators and 
$\hat n_{i\sigma} = c^+_{i\sigma} c_{i\sigma}$ is the electron number operator. 
Note that the above description of the electrons with charges $e$ includes a 
coupling to a magnetic field 
$\vec{B}(\vec{r}) = \vec{\nabla} \times \vec{A}(\vec{r})$, which is necessary 
for calculating the effects of current on the electronic states.

In what follows we shall work within Spin - Polarized - Hartree - Fock - 
Gorkov ($SPHFG$) approximation, which means that we have approximated the 
interaction term in Hamiltonian (\ref{Hamiltonian}) by the usual mean field
theory mapping:
$\hat n_{i\uparrow} \hat n_{i\downarrow} \rightarrow 
 \langle c_{i\downarrow} c_{i\uparrow} \rangle 
 c^+_{i\uparrow} c^+_{i\downarrow} + 
 \langle c^+_{i\uparrow} c^+_{i\downarrow} \rangle 
 c_{i\downarrow} c_{i\uparrow}$.
For the magnetic field we have chosen the Landau gauge where 
$\vec{B}=(0,0,B_z(x))$ 
and hence $\vec{A}=(0,A_y(x),0)$. Furthermore, we have assumed that the 
effective $SPHFG$ Hamiltonian is periodic in the direction parallel to the 
interface and therefore we work in $\vec{k}$ space in the $y$ direction but in 
real space in the $x$-direction (see Fig.\ref{Fig1}). Labelling the `planes`
(lines) in Fig.\ref{Fig1} by integers $n$ and $m$ at each $k_y$ point of the 
Brillouin zone we shall solve the following $SPHFG$ layer index ($n$) matrix 
equation for the retarded Green's function matrix $\hat G_{m'm}(\omega,k_y)$:
\begin{eqnarray}
 \sum_{m' k_y} \left(\omega \hat \tau_0 \delta_{n m'} - 
                     \hat H_{nm'}(k_y) \right) 
 \hat G_{m'm}(\omega,k_y) =
 \delta_{nm} 
 \label{HFG}
\end{eqnarray}
where $\hat \tau_0$ is the Pauli unit matrix and $\hat H_{nm}(k_y)$ is of the 
form:
\begin{widetext}
 \begin{eqnarray}
  \hat H_{nm}(k_y) = 
  \left(
  \begin{array}{cccc}
   \frac{1}{2} E_{ex}\delta_{nm} - T_- & - \Delta_n \delta_{nm} 
   & 0 & 0 \\
   - \Delta_n \delta_{nm} & \frac{1}{2} E_{ex} \delta_{nm} + T_+ 
   & 0 & 0 \\
   0 & 0 & - \frac{1}{2} E_{ex} \delta_{nm} - T_- 
   & \Delta_n \delta_{nm} \\
   0 & 0 & \Delta_n \delta_{nm} 
   & - \frac{1}{2} E_{ex} \delta_{nm} + T_+ 
  \end{array}
  \right) 
  \label{HHFG}
 \end{eqnarray}
\end{widetext}
with $T_{\pm} = (t cos(k_y \pm eA_y(n)) + \mu)\delta_{nm} + t \delta_{n,n+1}$. 

{\center
\vspace{0.5cm}
\subsection{\label{subsec2.b} Semi - infinite superconductor}
}

Our system consists of infinite number of layers ($n$), because of fact that we
have a semi-infinite superconductor. The $FM$ region spreads from $n = - N_F$ 
to $0$, while superconductor is defined for $n \geq 1$ (see Fig. \ref{Fig1}).
In order to describe semi-infinite system, we need infinite range matrix with
block elements 
$\hat M_{nm}(\omega,k_y) = \omega \hat \tau_0 \delta_{nm} - \hat H_{nm}(k_y)$ 
($n$, $m$ going from $- N_F$ through $0$ to $\infty$). In short, strictly 
speaking, we have to invert the infinite range matrix:
\begin{widetext}
 \begin{eqnarray}
  \hat M(\omega,k_y) = 
  \left(
  \begin{array}{ccccccccc}
   \hat M_{-N_F,-N_F} & \hat M_{-N_F,-N_F+1} & 0 & 0 & 0 & 0 & 0 & 0 & 0 \\
   \hat M_{-N_F+1,-N_F} & \hat M_{-N_F+1,-N_F+1} & \hat M_{-N_F+1,-N_F+2} & 
   0 & 0 & 0 & 0 & 0 & 0 \\
   0 & \ddots & \ddots & \ddots & 0 & 0 & 0 & 0 & 0 \\
   0 & 0 & \hat M_{0,-1} & \hat M_{00} & \hat M_{01} & 0 & 0 & 0 & 0 \\
   0 & 0 & 0 & \hat M_{1,0} & \hat M_{1,1} & \hat M_{1,2} & 0 & 0 & 0 \\
   0 & 0 & 0 & 0 & \ddots & \ddots & \ddots & 0 & 0\\
   0 & 0 & 0 & 0 & 0 & \hat M_{N_S,N_S-1} & \hat M_{N_S,N_S} & 
   \hat M_{N_S,N_S+1} & 0 \\
   0 & 0 & 0 & 0 & 0 & 0 & \ddots & \ddots & \ddots
  \end{array}
  \right) 
  \label{Hinf}
 \end{eqnarray}
\end{widetext}
To render the problem traceable, it is useful to define the surface Green's 
function ($SGF$) \cite{Turek} as
\begin{eqnarray}
 G^{sf}_{n,n}(\omega,k_y) = \{[\hat M_{n > N_S}(\omega,k_y)]^{-1}\}_{n,n}
 \label{SGFdef}
\end{eqnarray}
$\hat M_{n > N_S}(\omega,k_y)$ being semi-infinite submatrix for $n,m > N_S$ 
(lower right corner of the matrix in Eq.(\ref{Hinf})). The physical meaning of 
the $SGF$ is that it represents a square diagonal subblock corresponding to the 
semi-infinite bulk shown with a shaded region in the Fig. \ref{Fig1}.

Now we can assume, that electronic properties in the $SC$ differ from that in 
the bulk only over the finite distance from the interface ($1 < n < N_S$). 
$N_S$ can correspond to the distance of a few $SC$ coherence lengths. For 
$n > N_S$ the electronic properties are that of the bulk superconductor. In 
other words, we assume that the system for $n > N_S$ is homogeneous. Namely, 
for $n > N_S$ we have:
\begin{eqnarray}
 \begin{array}{c}
  \hat M_{n,n}(\omega,k_y) = \hat M_{N_S,N_S}(\omega,k_y) \\
  \hat M_{n,n+1}(\omega,k_y) = \hat M_{N_S,N_S+1}(\omega,k_y) \\
  \hat M_{n+1,n}(\omega,k_y) = \hat M_{N_S+1,N_S}(\omega,k_y)
 \end{array}
 \label{Hconst}
\end{eqnarray}
As a consequence the surface Green's function (\ref{SGFdef}) is 
also $n$-independent and can be obtained from the bulk $GF$ \cite{Turek}:
\begin{eqnarray}
 \hat G^{sf}(\omega,k_y) = \hat G_{N_S,N_S}(\omega,k_y) \times
 \nonumber \\
 (1 + \hat M_{N_S+1,N_S}(\omega,k_y) \hat G_{N_S,N_S+1}(\omega,k_y))^{-1}
 \label{SGFbulk}
\end{eqnarray}
where $\hat G_{n,m}(\omega,k_y)$ is the Fourier transform of the Green's 
function of the homogeneous superconductor $\hat G(\omega,k_x,k_y)$, namely 
$\hat G_{n,m}(\omega,k_y) = \frac{1}{N_{kx}} \sum_{k_x} 
 \hat G(\omega,k_x,k_y) e^{-i k_x (R_n-R_m)}$. Clearly, the Green's function 
$\hat G(\omega,k_x,k_y)$ for the case where the vector potential is equal to 
zero is given by the usual result:
\begin{widetext}
\begin{eqnarray}
 \hat G(\omega,k_x,k_y) = \frac{1}{\omega^2 - (\xi^2_{\bf k} + \Delta^2)} 
 \left(
 \begin{array}{cccc}
 \omega + \xi_{\bf k} & -\Delta & 0 & 0 \\
 -\Delta & \omega - \xi_{\bf k} & 0 & 0 \\
 0 & 0 & \omega + \xi_{\bf k} & \Delta \\
 0 & 0 & \Delta & \omega - \xi_{\bf k}
 \end{array}
 \right)
 \label{GFSCbulk}
\end{eqnarray}
\end{widetext}
where $\xi_{\bf k} = -2 t (cos(k_x) + cos(k_y)) - \mu$ is the single electron
energy spectrum.

The homogeneity of the system for $n > N_S$ leads to the conclusion, that 
the inversion of a semi-infinite block tridiagonal matrix (\ref{Hinf}) can be 
reduced to the inversion of the finite tridiagonal matrix:
\begin{widetext}
 \begin{eqnarray}
  \hat M(\omega,k_y) = 
  \left(
  \begin{array}{ccccccc}
   \hat M_{-N_F,-N_F} & \hat M_{-N_F,-N_F+1} & 0 & 0 & 0 & 0 & 0 \\
   \hat M_{-N_F+1,-N_F} & \hat M_{-N_F+1,-N_F+1} & \hat M_{-N_F+1,-N_F+2} & 
   0 & 0 & 0 & 0 \\
   0 & \ddots & \ddots & \ddots & 0 & 0 & 0 \\
   0 & 0 & \hat M_{0,-1} & \hat M_{00} & \hat M_{01} & 0 & 0 \\
   0 & 0 & 0 & \hat M_{1,0} & \hat M_{1,1} & \hat M_{1,2} & 0 \\
   0 & 0 & 0 & 0 & \ddots & \ddots & \ddots \\
   0 & 0 & 0 & 0 & 0 & \hat M_{N_S,N_S-1} & 
   \hat M_{N_S,N_S} - \hat \Gamma_{N_S}
  \end{array}
  \right) 
  \label{Hfin}
 \end{eqnarray}
\end{widetext}
In the above equation the effect of the semi-infinite homogeneous system is 
contained in the 'embedding potential' $\hat \Gamma_{N_S}$, which is related 
to the surface Green's function through:
\begin{eqnarray}
 \hat \Gamma_{N_S}(\omega,k_y) = \hat M_{N_S,N_S+1}(\omega,k_y) \times
 \nonumber \\
 \hat G^{sf}(\omega,k_y) \hat M_{N_S+1,N_S}(\omega,k_y)
 \label{Gamma}
\end{eqnarray}

Evidently, by increasing $N_S$ until $\hat G_{nm}(\omega,k_y)$ in the surface
region does not change would constitute an exact albeit numerical solution of
the Eq. (\ref{HFG}). All calculations reported here were performed for 
$N_S = 20$. However, we have checked that larger $N_S$ has no influence on the 
results.

{\center
\vspace{0.5cm}
\subsection{\label{subsec2.c} Finite temperature method}
}

The calculation of the various physical quantities, like particle 
concentration $n_n$, spin polarization $m_n$, order parameter $\Delta_n$, 
etc., needs the evaluation of integrals of the product of the Green's function 
and Fermi distribution function as in $\int d\omega G(\omega) f(\omega)$. 
While working on the real energy axis, we need a large number of integration 
points to get good accuracy, unless there are no singularities in the density 
of states. However, in our calculations, there is a Van Hove singularity in 
the $2D$ $DOS$ as well as $BCS$-like singularities. In fact we need 
approximately $10^3$ - $10^4$ points to get good accuracy, which is very 
time-consuming in self-consistent calculations. The situation is similar for 
summation over the Matsubara energies $\omega_{\nu} = (2 \nu +1) \pi i/\beta$ 
because of the poles of the Fermi function. There is an infinite number of 
poles, and infinite sum does not always converge well, so numerical treatment 
needs a large number of data points.

To overcome these difficulties we follow Ref. \cite{LitakMiller} and 
approximate the Fermi distribution function by:
\begin{eqnarray}
 f^c(\omega) = \frac{1}{\left(\frac{\omega+\sigma}{\sigma}\right)^{2N} + 1}
 \label{Fermic}
\end{eqnarray}
To get the correct behaviour near the Fermi energy ($E_F = 0$) we choose 
$2 N = \beta \sigma$. One can check, that the discrepancy between the
approximate function and the true one, starts to play role for energies lower
than $2 \sigma$. If we choose $2 \sigma > W$, where $W = 2 t$ is the half of 
the bandwidth, the approximation works very well. For example, for the
temperature $T = 10^{-2} t$ it is sufficient to consider only $150$ points to 
get very good accuracy. 

The poles of the approximate function lie on a circle given by
\begin{eqnarray}
 \left(\frac{\omega+\sigma}{\sigma}\right)^{2N} + 1 = 0
 \label{polescircle}
\end{eqnarray}
The solution of the above equation gives the poles in the form:
\begin{eqnarray}
 \omega = -\sigma + \sigma e^{(2 \nu + 1) \pi i / 2 N}
 \label{polesenergy}
\end{eqnarray}
In the limit $\sigma \rightarrow \infty$, the poles (\ref{polesenergy}) move to
the Matsubara energies, and the number of poles, $2 N$, becomes infinite.

The calculation of any physical quantity becomes:
\begin{eqnarray}
 \langle \hat O \rangle = \frac{2}{\pi} \sum^{2N-1}_{\nu = 0} 
 {\rm Re} G(\omega_{\nu}) e^{(2 \nu + 1) \pi i / 2 N}
 \label{Oaverage}
\end{eqnarray}
where $\omega_{\nu}$ is given by Eq.(\ref{polesenergy}).

{\center
\vspace{0.5cm}
\subsection{\label{subsec2.d} Self-consistency with the Ampere's law}
}

As usual in problems of interfaces, a large number of equations has to be 
solved self-consistently. In our case the parameters, to be determined 
self-consistently at each layer $n$, are: the total electron density $n_n$, 
the spin polarization $m_n$, the $SC$ order parameter $\Delta_n$, the current 
parallel to the interface $J_y(n)$ and the vector potential $A_y(n)$. Most of 
these quantities can be expressed in terms of the corresponding 
layer - diagonal elements of the Green's functions $\hat G_{nn}(\omega,k_y)$ 
(see Eq.(\ref{HHFG})). The exception is the vector potential, for which we need 
an additional (Maxwell's) equation. These relations are 
(see Eq.(\ref{Oaverage})):
\begin{eqnarray}
 n_n = n_{n\uparrow} + n_{n\downarrow} = 
 \frac{2}{\beta} 
 \sum_{ky} \sum^{2N-1}_{\nu = 0}
 \times
 \nonumber\\
 {\rm Re} \left\{
 (G^{11}_{nn}(\omega_{\nu},k_y) + G^{33}_{nn}(\omega_{\nu},k_y))
  e^{(2 \nu + 1) \pi i / 2 N}
  \right\}
 \label{n}
\end{eqnarray}
\begin{eqnarray}
 m_n = n_{n\uparrow} - n_{n\downarrow} = 
 \frac{2}{\beta} 
 \sum_{ky} \sum^{2N-1}_{\nu = 0}
 \times
 \nonumber\\
 {\rm Re} \left\{
 (G^{11}_{nn}(\omega_{\nu},k_y) - G^{33}_{nn}(\omega_{\nu},k_y))
 e^{(2 \nu + 1) \pi i / 2 N}
 \right\}
 \label{m}
\end{eqnarray}
\begin{eqnarray}
 \Delta_n = U_n \sum_{k_y} 
 \langle c_{n\downarrow}(k_y) c_{n\uparrow}(k_y) \rangle = 
 \nonumber\\
 \frac{2 U_n}{\beta} 
 \sum_{ky} \sum^{2N-1}_{\nu = 0}
 {\rm Re} \left\{
 G^{12}_{nn}(\omega_{\nu},k_y) e^{(2 \nu + 1) \pi i / 2 N}
 \right\}
 \label{Delta}
\end{eqnarray}
where $G^{ij}_{nn}(\omega,k_y)$ is the $ij$-th element of the matrix Green's 
function, which is the solution of the $SPHFG$ equation (Eq. (\ref{HFG})).

The current, for spin up electrons, in the $y$-direction can be calculated 
from the relation:
\begin{eqnarray}
 J_{y\uparrow (\downarrow)}(n) = 
 \frac{4 e t}{\beta} \sum_{k_y} sin(k_y - e A_y(n)) 
 \times
 \nonumber\\
 \sum^{2N-1}_{\nu = 0}
 {\rm Re} \left\{
 G^{11(33)}_{nn}(\omega_{\nu},k_y) e^{(2 \nu + 1) \pi i / 2 N}
 \right\}
 \label{current}
\end{eqnarray}
which follows from the continuity equation for the charge 
($e \frac{d n_i}{d t} = - e [n_i, H]_- = - \sum_j J_{ij}$).

This current will give rise to a vector potential $A_{new}(\vec{r})$ which 
will have to be used to update $A(\vec{r})$ in Eq.(\ref{HFG}) at the end of 
each self-consistency cycle. We calculated this new vector potential by solving 
numerically Ampere's law, $\frac{d^2 A_y(x)}{d x^2} = - 4 \pi J_y(x)$, which 
for the lattice problem at hand, is 
\begin{eqnarray}
 A_y(n+1) - 2 A_y(n) + A_y(n-1) = - 4 \pi J_y(n)
 \label{Maxwell}
\end{eqnarray}

In all our calculations the Eqs.(\ref{n})-(\ref{Maxwell}) have been solved 
in each iteration step until self-consistency has been achieved. The 
numerical calculations have been performed at temperature $T = 10^{-2} t$ for 
$150$ energy points, $120$ - $k_y$ points and $40$ layers ($20$ ferromagnetic 
and $20$ superconducting). Typically self-consistency, at the level of
$0.01 \; \%$ on all densities in Eqs.(\ref{n})-(\ref{Maxwell}), has been 
achieved after $100$ - $200$ iterations.

\vspace{1cm}
%%%%%%%%%%%%%%%%%%%%%%%%%%%%%%%%%%%%%%%%%%%%%%%%%%%%%%%%%%%%%%%%%%%%%%%%%%%%%%

{\center
\section{\label{sec3} ANDREEV BOUND STATES IN FERROMAGNET}
}

{\center
\vspace{0.5cm}
\subsection{\label{subsec3.a} Superconducting and ferromagnetic order 
            parameters}
}

Since we have determined the ferromagnetic (\ref{m}) and superconducting 
(\ref{Delta}) order parameters on both sides of the interface fully 
self-consistently, we were able to study both $SC$ and $FM$ proximity effects. 
This means, that we were able to describe situations in which ferromagnetism 
and superconductivity coexist near the interface.

\begin{figure}[h]
 \resizebox{8cm}{5cm}{
  \includegraphics{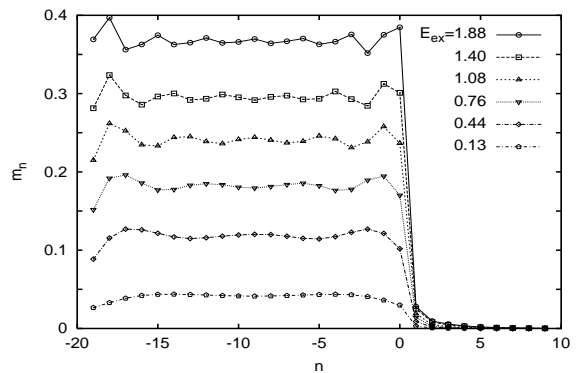}}
 \caption{\label{Fig2} Magnetization as a function of the distance from the 
 interface for a number of exchange fields ($E_{ex}$) and $U_S = -2$, which
 gives $\Delta_S = 0.376$ in units of the hopping integral $t$.}
\end{figure}

Firstly, we want to discuss the magnetic proximity effect, namely the entering 
of the spin polarization into the superconductor. The typical behaviour of the 
spin polarization ($m_n = n_{n\uparrow} - n_{n\downarrow}$) is plotted in the 
Fig. \ref{Fig2} as a function of the layer index $n$ for a number of exchange 
splittings $E_{ex}$. 
We can see that the spin polarization exponentially decays over the distance of 
the $SC$ coherence lengths $\xi_S$, which in the present case is 
$\xi_S \approx 3$ in units of the lattice constant $a$. We have checked this by 
explicit calculations of $\xi_S$ for various values of the superconducting 
energy gap. So we can conclude that effect of proximity of the ferromagnet on 
the superconductor is very similar to case of the Meissner effect, where an 
external magnetic field is excluded from the sample. Note however that while 
the magnetic field is excluded on the spatial scale of the penetration depth 
the effective exchange field drops to zero within the distance of the 
coherence length $\xi_S$ in the superconductor.

More interesting is the behaviour of the spin polarization on the ferromagnetic 
side of the heterostructure. As it is seen from Fig. \ref{Fig2}, it oscillates 
around its bulk value with period depending on the exchange splitting $E_{ex}$ 
but not on the superconducting gap. However we were not able to fit the 
numerical points to any simple relation between parameters of the model. The 
best fit we have is $m_n \propto sin{((1.74 E_{ex} + 0.5) n)}$. Whilst this 
means that the period of the oscillations is a linear function of the exchange 
splitting the above relation is valid only for the thickness of the $FM$ 
slab $d = 20$ layers. For different thicknesses we have different relations and 
different numbers of period within the slab. For example, for $d=10$, $m_n$ 
shows one period less for the corresponding exchange splittings $E_{ex}$.

We now turn to the superconducting proximity effect, namely the leakage of 
the $SC$ properties into the non-superconducting region. In our case, although 
$\Delta_n = 0$ on the ferromagnetic side, due to the fact that for $n \leq 0$, 
$U_n = 0$, the pairing amplitude 
$\chi_n = \langle c_{n\downarrow} c_{n\uparrow} \rangle$ 
can, in general, be non-zero and it usually is.

The proximity effect is well understood in the case of non-magnetic 
metal - superconductor interface ($NM$/$SC$) \cite{LambertRaimondi} in terms of 
the Andreev reflections \cite{Andreev}. The Andreev process describes a 
situation in which an electron impinging onto the $NM$/$SC$ interface is 
reflected back as a hole (with opposite spin) and a Cooper pair is created in
the $SC$. Moreover, the movement of the electrons and holes in the $NM$ close 
to the interface (over the distance of the $SC$ coherence length $\xi_S$) is 
highly correlated. The pairing amplitude, in this case, enters the normal metal 
and decays exponentially over the $\xi_S$.

\begin{figure}[h]
 \resizebox{8cm}{5cm}{
  \includegraphics{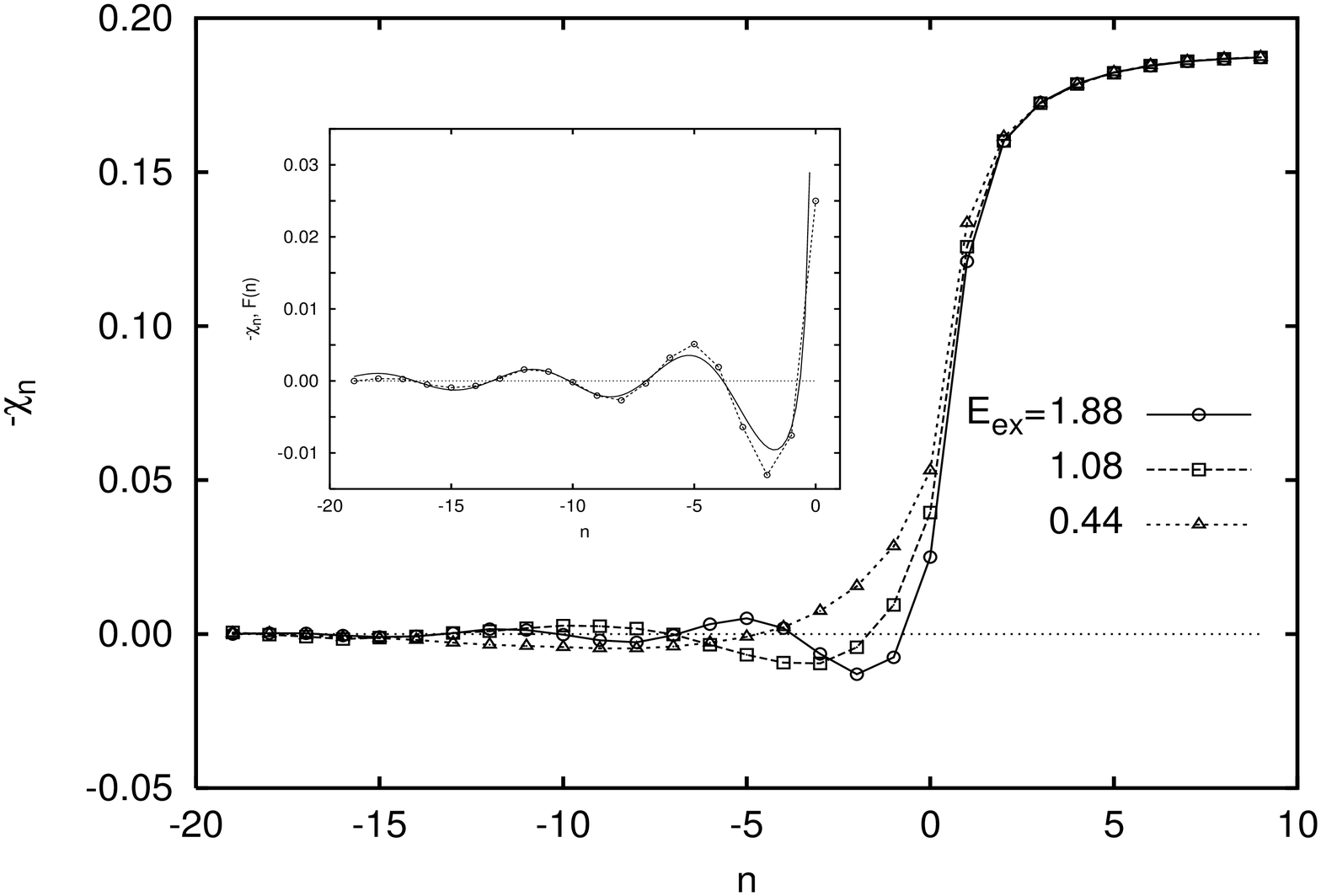}}
 \caption{\label{Fig3} The pairing amplitude $-\chi_n$ vs layer index $n$ for
 three values of the exchange field $E_{ex}$. Note that we plot $-\chi_n$ which
 corresponds to positive $\Delta_n$ for $U_n > 0$. Inset: Comparison of the
 numerical results with the analytical formula 
 $\chi_n \propto sin{(n/\xi_F)}/(n/\xi_F)$ for the exchange field 
 $E_{ex}=1.88$, which gives $\xi_F=1.06$. In fact we had to change $\xi_F$ by 
 $5\%$ in order to get better agreement. The solid line is our fit: 
 $\chi_n = 0.021 sin{(n/\tilde\xi_F - 2.5)}/(n/\tilde\xi_F)$, where 
 $\tilde\xi_F = \xi_F/1.05$.}
\end{figure}

When the normal metal is replaced by a ferromagnet one might naively
expect that the proximity effect will be suppressed due to the pair breaking
effect of the exchange field \cite{deJong} . This suggests that the pairing 
amplitude $\chi_n$ will decay over a distance much shorter than the 
superconducting coherence length. Remarkably, as we can see in the 
Fig. \ref{Fig3}, this is not the case. 
Instead, we can observe a very long range proximity effect with an oscillating 
pairing amplitude. This effect has been first noted by A. I. Buzdin 
{\it et al.} \cite{Buzdin82,Buzdin90} and have been studied in several recent
publications \cite{Radovic,Demler,Vecino,Halterman,Bagrets}. According to them,
it can be attributed to a $FFLO$ - like phenomena 
\cite{FuldeFerrell, LarkinOvchinnikov} in $FM$/$SC$ heterostructures. 
The period of the oscillations of the $\chi_n$ depends on the exchange 
splitting $E_{ex}$ (see Fig. \ref{Fig3}) and, similarly to the oscillations of 
$m_n$, does not depend on the superconducting energy gap. Clearly this 
behaviour is different from $NM$/$SC$ case, where the proximity effect depends 
on the $SC$ coherence length $\xi_S$. Here, it depends only on the properties 
of the ferromagnet. Moreover, the numerically obtained behaviour of the pairing 
amplitude $\chi_n$ is consistent with the analytical formula
\begin{eqnarray}
 \chi_n \propto sin{(n/\xi_F)}/(n/\xi_F)
 \label{chi_analit}
\end{eqnarray}
with $\xi_F = 2 t / E_{ex}$ being ferromagnetic coherence length. Reassuringly
this result is fully consistent with those of Ref. 
\cite{Demler,Vecino,Halterman,Bagrets}. The comparison our numerical results 
with the formula (\ref{chi_analit}) for the exchange splitting $E_{ex} = 1.88$, 
which gives $\xi_F = 1.06$ is depicted in the inset of the Fig. \ref{Fig3}. In 
fact we had to change $\xi_F$ by $5\%$, in order to get better fit. So the 
solid line in the inset of the Fig. \ref{Fig3} corresponds to the formula 
$\chi_n = 0.021 sin{(n/\tilde\xi_F - 2.5)}/(n/\tilde\xi_F)$ with 
$\tilde \xi_F = \xi_F/1.05$. It turns out that as exchange splitting becomes 
smaller the fitting to the analytical formula (\ref{chi_analit}) becomes 
better. The small discrepancy in the coherence length may come from the fact, 
that we have used bulk value of the exchange field while calculating the 
coherence length. However in our system, the spin polarization oscillates 
around its bulk value, and if we calculate the average value of the spin 
polarization $\bar{m} = \frac{1}{N_F+1}\sum_{n \leq 0} m_n$, we get a little 
bit smaller value than the bulk one. So, we conclude, that the effective 
exchange splitting $\tilde E_{ex}$ also has to be smaller than its bulk value. 
Indeed, when we used $\tilde E_{ex}$ to calculate the coherence length, the 
agreement is improved. Another reason may be the approximate form of the 
formula for the coherence length in the lattice case 
$\xi_F \approx 2 t / E_{ex}$.

{\center
\vspace{0.5cm}
\subsection{\label{subsec3.b} Density of states and the Andreev levels}
}

The proximity effect manifests itself not only in the pairing amplitude and the 
spin polarization characteristics but also in the density of states ($DOS$). 
The layer resolved $DOS$ is defined as
\begin{eqnarray} 
 \rho_n(\omega) \def \rho_{n\uparrow}(\omega) + \rho_{n\downarrow}(\omega) = 
 \nonumber \\
 -\frac{1}{\pi}\sum_{k_y} {\rm Im} 
 (G^{11}_{nn}(\omega + i0^+,k_y) + G^{33}_{nn}(\omega + i0^+,k_y))
 \label{DOS_n}
\end{eqnarray}
with $G^{ij}_{nn}(\omega + i0^+,k_y)$ being the $ij$-th element of the matrix 
Green's function - solution of the Eq. (\ref{HFG}). We shall also use the 
density of states integrated over the total sample: 
$\rho_{tot}(\omega) = \sum_n \rho_n(\omega)$.

At this point we want to stress that, when we calculate thermodynamic
quantities (Eq.(\ref{n})-({\ref{current}})) we work on the complex energy 
plane, but calculating the $DOS$ we solve Eq. \ref{HFG} on the real energy axis 
only once. In fact we have added a small imaginary part ($0^+ = 0.005$) for 
numerical purposes. The quantities: $n_n$, $m_n$, $\Delta_n$, $A_y(n)$ as 
determined by the previous self-consistent procedure allow us to find the 
Green function on the real energy axis in only one step. We also check if 
the new quantities like $n_n$, $m_n$, etc. agree with the old ones. For all 
such calculations the agreement has been found to be excellent. 

While the general solution of Eq.(\ref{HFG}) allows for a current flow, in this 
section we show the results for a solution for which the current has been 
constrained to be zero.

To appreciate the effect of spin polarization in the $FM$ slab, let us recall
again the case of no exchange splitting. If the non-magnetic metal ($NM$) is in 
the contact with superconductor ($SC$), the density of states on the $NM$ side 
shows features of the $SC$ $DOS$, i.e. the pseudo-gap opens up around the Fermi 
energy in the $DOS$. However it never turns into a true gap, namely the $NM$ 
$DOS$ always possess low energy excitations. Moreover, the gap in the $SC$ 
$DOS$ also starts to fill up, due to the proximity of the normal metal. So in 
this case we deal with gapless superconductivity close to interface of the 
$NM$/$SC$ heterostructure.

\begin{figure}[h]
 \resizebox{8cm}{!}{
  \includegraphics{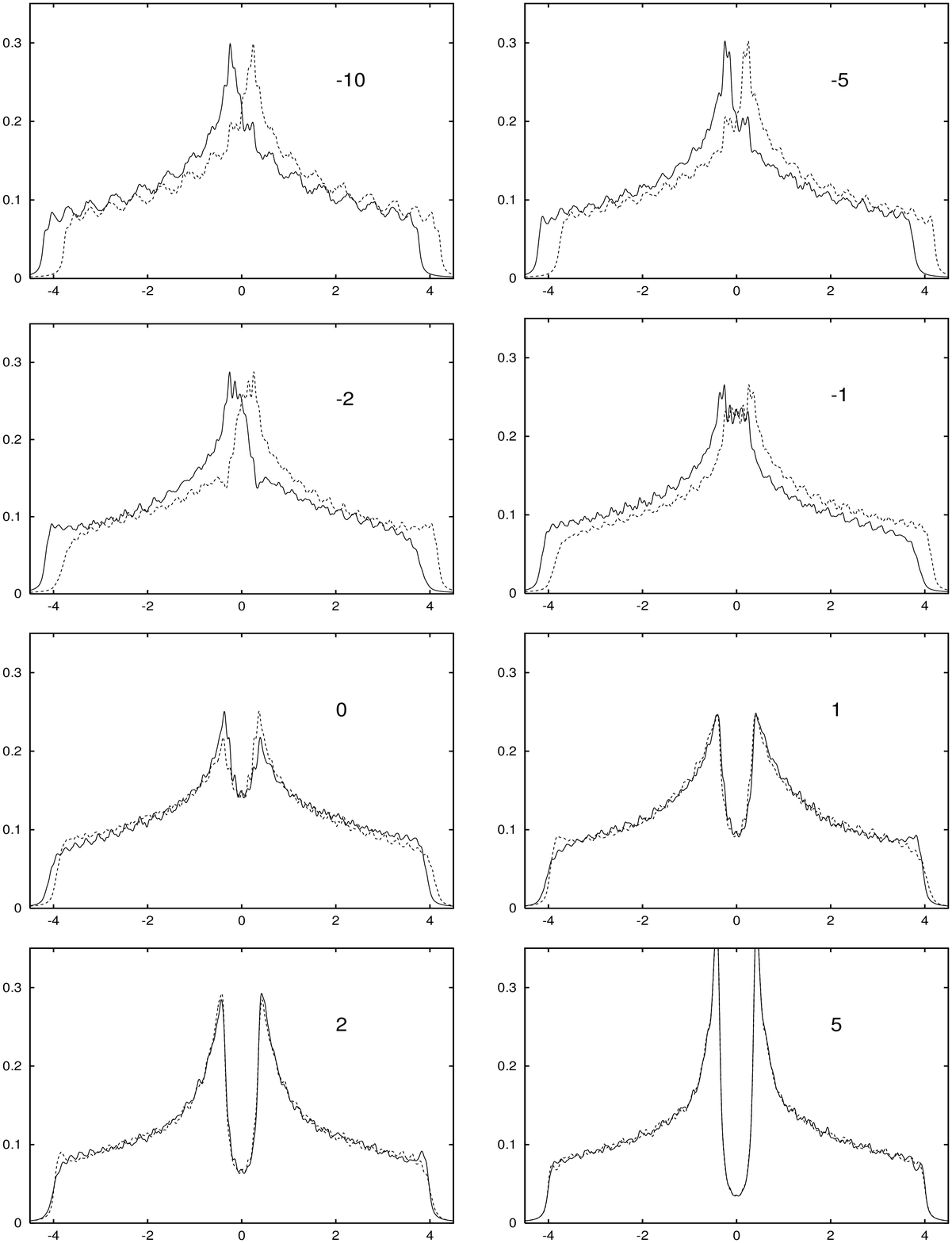}}
 \caption{\label{Fig4} The layer resolved density of states for the spin up 
 (solid line) and spin down (dashed line) electrons for the exchange splitting 
 $E_{ex}=0.44$. The layer index is shown in the panels.}
\end{figure}

Compared to the above summary, the ferromagnet - superconductor interface is
enriched with a number of new features. For instance, the $SC$ $DOS$ splits, 
if the exchange field is small. For strong enough exchange field, $SC$ gap can 
be suppressed, indicating transition to the non-$SC$ (normal) phase. An example 
of the layer resolved density of states for spin up (solid line) and down 
(dashed) is shown in the Fig. \ref{Fig4}.
Evidently there is a pseudogap in the $FM$ $DOS$, indicating that 
superconducting correlations are present in the $FM$ layers. This effect is 
particularly strong for the layer index $n=0$ and $-1$. On the other hand, the
proximity of the ferromagnet leads to existence of the low energy states in 
the $SC$ $DOS$. This effect is also observed in $NM$/$SC$ heterostructure but
here we also notice differences in the spin up and spin down $DOS$ on the $SC$ 
layers close to the interface. These are clear manifestations of the $FM$ 
proximity effect.

\begin{figure}[h]
 \resizebox{8cm}{5cm}{
  \includegraphics{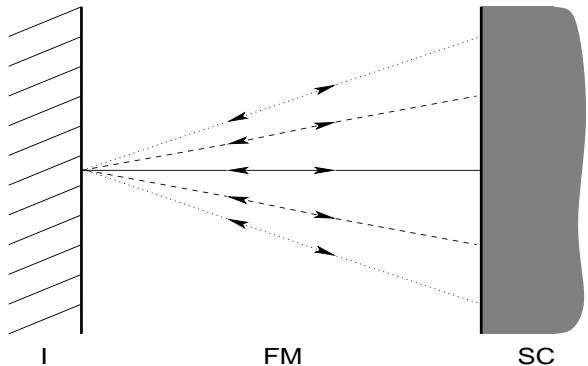}}
 \caption{\label{Fig5} Schematic view of the semi-classical paths of impinging
 electrons and reflected holes corresponding to the different Andreev bound 
 states.}
\end{figure}

The most interesting physics of the heterostructures we consider, is the
formation of Andreev bound states in the $NM$ or $FM$ layers. They are
`particle in a box` like states which can be associated with the semiclassical
orbits bouncing back and forth between the $SC$ and $I$ regions, as depicted in
the Fig. \ref{Fig5}. As is well known in the case of $NM$ layers the reflections
at the $SC$ are Andreev ones, while those at the $I$ are normal reflections
\cite{deGennes-SJames}. The arrows in the Fig. \ref{Fig5} point along the
momenta of the particle segment of the orbit. Its $y$ component is $k_y$ which
labels the state. Clearly, the collection of states labelled in this way forms 
an Andreev band.

\begin{figure}[h]
 \resizebox{8cm}{5cm}{
  \includegraphics{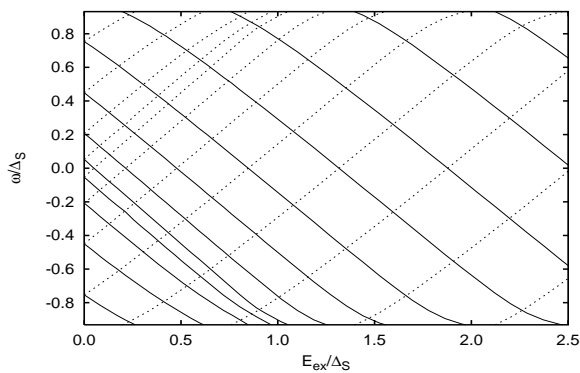}}
 \caption{\label{Fig6} The position of the Andreev bound states vs exchange
 splitting $E_{ex}$ for the $20$ layers thick ferromagnet. The solid (dashed) 
 line corresponds to the spin up (down) electrons. Note that energy is now 
 measured in units of the bulk superconductor order parameter $\Delta_S$. The 
 basic physics represented by this plot is the same as the 
 $\omega/\Delta_S$ vs $d$ (thickness of the $FM$ sample) curves of Ref.
 \cite{Vecino}.}
\end{figure}

For the normal metal - superconductor heterostructure, Andreev bound states
(bands) are symmetrically placed with respect to the Fermi energy. Of course 
the position of these states changes with the thickness of the sample. As
we increase the size of the sample, the Andreev bound states approach the Fermi
energy, but they will never reach it. In other words, there is no possibility 
of crossing the Fermi level. Somewhat surprisingly, the situation is quite 
different for the $FM$/$SC$ structure \cite{Vecino}. In this case the bound
state energies can cross the Fermi level. In short they can become the zero -
energy mid - gap states \cite{Kashiwaya}. This situation is illustrated, for 
$k_y = 0$, in the Fig. \ref{Fig6}, where the bound state energies are plotted
against the exchange splitting.

\begin{figure}[h]
 \resizebox{8cm}{5cm}{
  \includegraphics{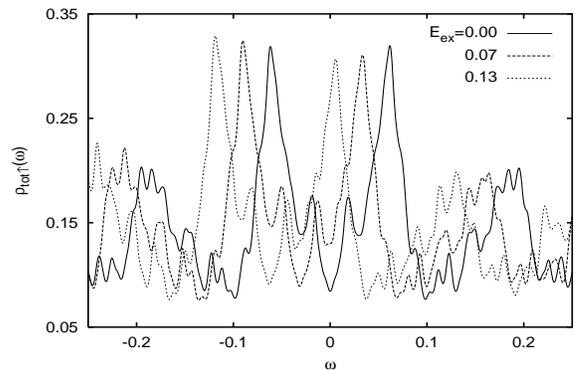}}
 \caption{\label{Fig7} The total density of states 
 ($\rho_{tot\uparrow}(\omega) = \sum_n \rho_{n\uparrow}(\omega)$ of the spin 
 up electrons for various values of the exchange splitting $E_{ex}$ indicated on
 the picture. Note the shift of the Andreev bound states (bands) with exchange
 splitting. The $DOS$ for spin down electrons can be easily obtained by
 reflection of the spin up electrons $DOS$ with respect to the Fermi energy
 $\varepsilon_F = 0$.}
\end{figure}

If we take into account states for different $k_y$ (Andreev band), the 
situation is more complicated as in general the states with different 
symmetry of the wave function can mix, and so we are not able to separate 
Andreev bands so easily. The example of such Andreev bands rather than states 
is shown in the Fig. \ref{Fig7}, where the density of states of the spin up 
electrons is plotted for a few values of the exchange splitting. 

As we shall report presently the fact that the Andreev bound states can cross 
the Fermi level has remarkable consequences: it leads to the generation of 
the spontaneous current in the ground state.

In fact the situation is rather similar to that of exotic superconductors 
(with non-$s$-wave order parameter), where the surface $DOS$ shows the 
zero-energy state, leading to the spontaneously generated current. However the 
origin of this state is quite different, as there it comes from the symmetry 
related sign change of the order parameter at the surface \cite{Kashiwaya}. 
Here, in the $FM$/$SC$ heterostructure, the zero-energy state comes from the 
oscillatory behaviour of the pairing amplitude ($\chi$) in $FM$, which changes
its phase by $\pi$ each time crosses zero. Nevertheless it leads to the same 
effect: a spontaneously generated current.

\vspace{1cm}
%%%%%%%%%%%%%%%%%%%%%%%%%%%%%%%%%%%%%%%%%%%%%%%%%%%%%%%%%%%%%%%%%%%%%%%%%%%%%%

{\center
\section{\label{sec4} CURRENT IN THE GROUND STATE}
}

{\center
\vspace{0.5cm}
\subsection{\label{subsec4.a} Splitting of the zero energy state}
}

\begin{figure}[h]
 \resizebox{8cm}{5cm}{
  \includegraphics{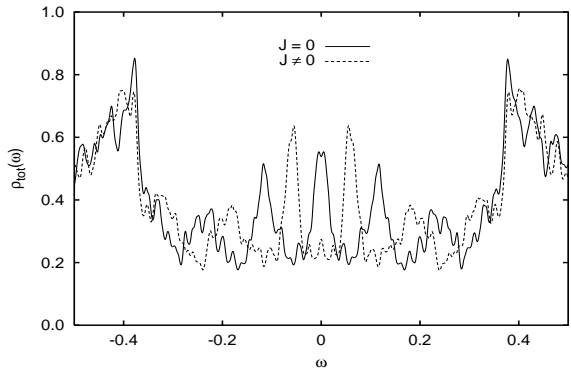}}
 \caption{\label{Fig8} The Doppler splitting of the zero energy state in the 
 density of states 
 $\rho_{tot}(\omega) = 
  \sum_n (\rho_{n\uparrow}(\omega) + \rho_{n\downarrow}(\omega))$ caused by the
  spontaneous current for $E_{ex}=0.13$. The solid (dashed) line corresponds to 
  the solution without (with) the current in the ground state.}
\end{figure}

The most remarkable feature of our calculations is that the solution of the 
$SPHFG$ equations frequently converges to a solution with the finite current 
$j_y(n)$ even though the external vector potential is zero. The generation of
the spontaneous current is strictly related to the crossing of the Andreev 
bound states through the Fermi energy. In the Fig. \ref{Fig7}, we have plotted
the $DOS$ for different values of the exchange splitting for calculations where
the current was allowed to develop in the $y$ - direction. For the 
$E_{ex}=0.13$ we observe that spontaneous current is generated in the ground 
state. We have checked that the solution with the current flowing is the true 
ground state, as it has energy lower than solution without the current. 
Physically, in the presence of the current, the zero energy states split 
because of the momentum dependence of a $\vec{p} \cdot \vec{v}$ contribution 
to the quasiparticle energies where the velocity vector is defined by the 
current $\vec{J} = e n \vec{v}$. The example of such a Doppler splitting in the 
$DOS$ is shown in Fig. \ref{Fig8} for the exchange field $E_{ex}=0.13$.
We can see that, besides the zero energy state, the other Andreev states also 
are split. It turns out that the splitting of these states is only weakly 
sensitive to the exchange field $E_{ex}$, but it strongly depends on the 
thickness of the ferromagnetic slab and it decreases as the thickness 
increases. 

Interestingly, we have found a simple relation between the splitting of the 
zero energy state $\delta$ and the vector potential $A_y(n)$. This is given by:
\begin{eqnarray}
 \delta \approx 2 e t \bar{A}_y
 \label{splitting}
\end{eqnarray}
where $\bar{A}_y$ is the value of the vector potential $A_y(n)$ averaged over 
the $FM$ side of the system only, i.e. 
$\bar{A}_y = \frac{1}{N_F+1}\sum_{n \leq 0} A_y(n)$. 

\begin{figure}[h]
 \resizebox{8cm}{5cm}{
  \includegraphics{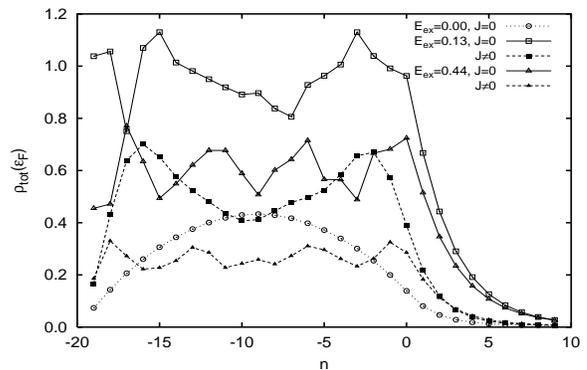}}
 \caption{\label{Fig9} The layer resolved density of states at the Fermi energy
 $\varepsilon_F = 0$ for different exchange splittings. The filled (empty)
 symbols correspond to the solution with (without) the current.}
\end{figure}

Corresponding to the splitting of the zero energy states, when the current 
flows, we find differences in the $DOS$ at the Fermi level. In fact, the 
difference is quite small. The main behaviour remains the same. The comparison 
of the density of states at the Fermi energy for both cases 
($J=0$ and $J \neq 0$) is shown in the Fig. \ref{Fig9}. 
The $DOS$ for the $NM$/$SC$ structure (dashed line with circles) is also drawn.
As we can see, the main features remain more or less the same, in particular 
the period of the oscillations. Similarly like the spin polarization 
(Fig. \ref{Fig2}) and pairing amplitude (Fig. \ref{Fig3}), the density of 
states at the Fermi level shows decreasing period of the oscillations as the 
exchange splitting increases.

{\center
\vspace{0.5cm}
\subsection{\label{subsec4.b} Spontaneous current and the magnetic flux}
}

\begin{figure}[h]
 \resizebox{8cm}{5cm}{
  \includegraphics{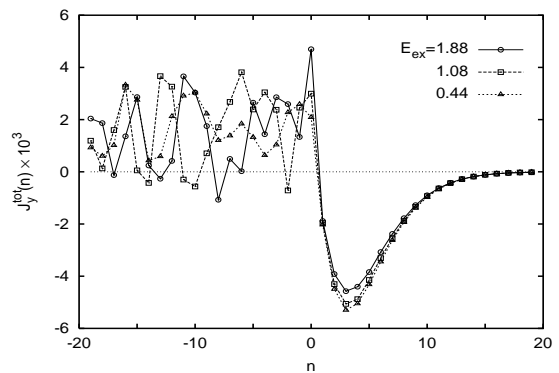}}
 \caption{\label{Fig10} The total (spontaneous) current
 $j^{tot}_y(n) = j_{y\uparrow}(n) + j_{y\downarrow}(n)$ flowing parallel to 
 the $FM$/$SC$ interface for a number of exchange splittings.}
\end{figure}

As we already discussed, the zero energy state is responsible for the 
generation of the spontaneous current in the system. The typical example of 
such a current, flowing parallel to the $FM$/$SC$ interface, 
($j^{tot}_y(n) = j_{y\uparrow}(n) + j_{y\downarrow}(n)$) is shown  in the 
Fig. \ref{Fig10} for a few values of the exchange splitting. 
Evidently, the behaviour of the current, as a function of the layer 
index, is very similar to the density of states at the Fermi level. For
example, if we compare the dashed curve with the triangles in Fig. \ref{Fig10} 
to the dashed or solid line with triangles in Fig. \ref{Fig9}, we readily 
notice the similarity in the oscillating nature of the current and $DOS$ 
respectively.

Another important issue is the distribution of the current through the whole
trilayer structure. We find that it flows mostly in the positive $y$ direction 
on ferromagnetic side and in the negative direction in the superconductor.
Notably the total current, integrated over the whole sample, is equal to zero 
within numerical accuracy. This is as it should be for the true ground state 
and found to be in the $FFLO$ state, where the current associated with the 
unpaired electrons is balanced by the supercurrent flowing in the opposite 
direction. 

While the total current in the system is equal to zero, the current integrated
over $FM$ side only ($J^{tot}_{FM} = \sum_{n \leq 0} J^{tot}_y(n)$)
has a finite value, almost independent of the exchange splitting. But that 
current is very sensitive to the thickness of the $FM$ slab, and for example 
for $N_F=10$ $J^{tot}_{FM} \approx 0.08$ ($0.008$ per layer), while for 
$N_F=20$ we have $J^{tot}_{FM} \approx 0.035$ ($0.00175$ per layer). This
suggests that effect associated with the spontaneous current is very important 
in the samples of the small thicknesses and it seems to play minor role if the 
size of $FM$ part of the system is large.

Obviously, the above current distribution should lead to the generation 
of the magnetic flux through the sample. In fact we have found that again, as 
the current, the magnetic flux bounded by the layers only weakly depends on the 
exchange splitting but it does change with the distance from the interface. For 
$N_F=10$ the magnetic flux $\Phi^{tot} = \sum_{n \leq 0} \Phi_n$ is nearly 
equal to the half of the flux quantum $\Phi_0$, 
$\Phi^{tot} \approx 0.45 \Phi_0$. This gives a flux per plaquette, associated 
with the layers $n$ and $n + 1$, $\Phi_n \approx 0.045 \Phi_0$. On the other 
hand, for $N_F=20$ we have $\Phi^{tot} \approx 0.25 \Phi_0$ and 
$\Phi_n \approx 0.0125 \Phi_0$ respectively.

\begin{figure}[h]
 \resizebox{8cm}{5cm}{
  \includegraphics{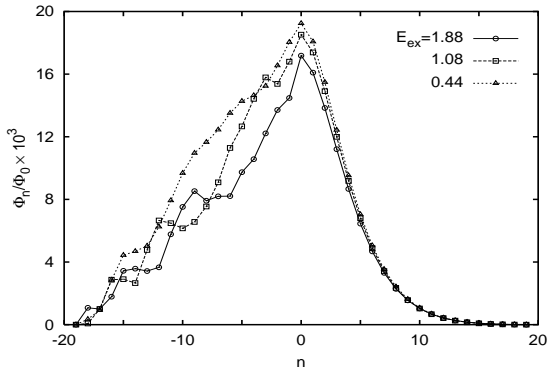}}
 \caption{\label{Fig11} The (spontaneous) magnetic flux per plaquette 
 associated with layer $n$ and $n+1$ for a number of exchange splittings. Each 
 curve corresponds to the related current shown in the Fig. \ref{Fig10}.}
\end{figure}

The typical layer dependence of the spontaneous magnetic flux $\Phi_n$ 
(magnetic flux through a plaquette) is shown in the Fig. \ref{Fig11}.
We can see that while there are small differences in the behaviour on the $FM$
and $SC$ sides of the sample, related to the different current distributions 
(see the Fig. \ref{Fig10}), the total magnetic flux $\Phi^{tot}$ is roughly the 
same and more or less independent of the exchange splitting.

{\center
\vspace{0.5cm}
\subsection{\label{subsec4.c} Quasiparticle current vs supercurrent.}
}

\begin{figure}[h]
 \resizebox{8cm}{5cm}{
  \includegraphics{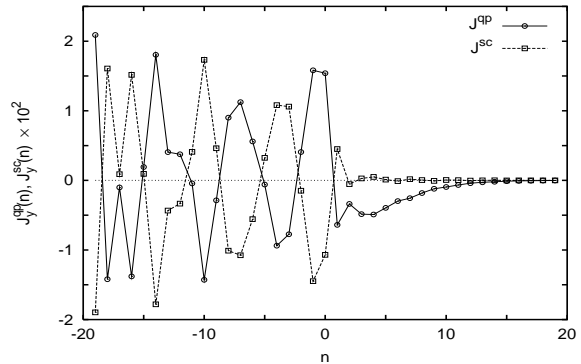}}
 \caption{\label{Fig12} Layered resolved quasiparticle current (solid line) 
          and supercurrent (dashed line) for exchange splitting $E_{ex}=1.88$. 
	  This corresponds to the total current (solid line with circles) shown 
	  in Fig. \ref{Fig10}.}
\end{figure}

As we have recalled above, in the bulk $FFLO$ state the total current vanishes. 
Moreover, the quasiparticle current $J^{qp}$ is exactly equal to the 
supercurrent $J^{sc}$ (related to the $G^{12}_{nn}(\omega, k_y)$ in 
Eq. (\ref{HFG})) with opposite sign. This means that in a bulk layered 
superconductor, these two currents cancel each other layer by layer, i.e.  
$J^{qp}(n) = - J^{sc}(n)$. Evidently this is the case because the exchange 
field and superconducting pairing potential are present in the whole system. 
By contrast, in the case of the $FM$/$SC$ heterostructure, the exchange field 
and the pairing potential are spatially separated. In this case $J^{qp}(n)$ is 
not equal to $- J^{sc}(n)$ within each layer, but it is when integrated over 
the whole sample, i. e. 
\begin{eqnarray}
 \sum_n J^{qp}(n) = - \sum_n J^{sc}(n)
 \label{Jqp-sc}
\end{eqnarray}
Thus the lack of exact cancellation of the current layer by layer leads to 
a finite current on both sides of the interface, but zero in the whole sample. 
An example of such behaviour is shown in Fig. \ref{Fig12}.
We see, that the quasiparticle part of the total current (solid line) is almost 
in anti-phase to the supercurrent (dashed line). Moreover, on the 
superconducting side, the current is carried mainly by quasiparticles. This is 
rather surprising since usually one expects current to be related to the 
Cooper pairs. However due to the proximity of the ferromagnet, there are also 
`normal` particles in the superconductor near the interface. These certainly 
contribute to the total current, like in the bulk $FFLO$ state. In our system, 
only the Cooper pairs close to the interface can feel the exchange field of the 
ferromagnet and have moving center of mass. Note that magnetization goes to
zero very quickly as we go away from the interface (see Fig. \ref{Fig2}).
Clearly this makes the supercurrent equal to zero except within a few $SC$ 
layers.

Interestingly we also found that whenever the pairing amplitude $\chi_n$ 
changes sign, the local quasiparticle current as well as the local supercurrent 
becomes large. In other words, the maxima of the $J^{qp}(n)$ and $J^{sc}(n)$ 
correspond to the zeros of the pairing amplitude $\chi_n$. 

{\center
\vspace{0.5cm}
\subsection{\label{subsec4.d} The pairing amplitude at I/FM interface}
}

\begin{figure}[h]
 \resizebox{8cm}{5cm}{
  \includegraphics{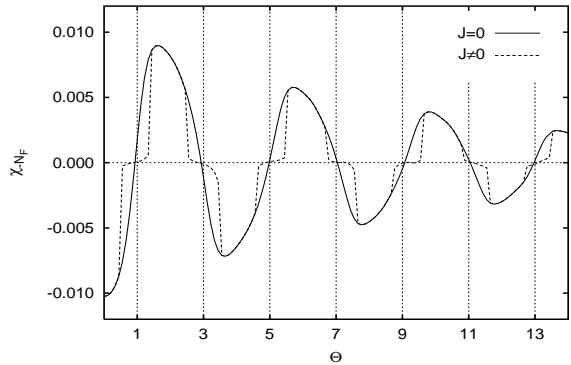}}
 \caption{\label{Fig13} Pairing amplitude at the $FM$ surface opposite to
 superconductor ($n = - N_F$) as a function of the dimensionless parameter 
 $\Theta = 2.79 d E_{ex}/\pi t$.}
\end{figure}

As we have mentioned above, the ground state does not carry current for all 
values of the exchange splitting. We have shown, that whenever there is a 
current flowing, there are the Andreev bound states crossing the Fermi level. 
Also we have found another quantity which is related to the current flowing. 
Namely, the pairing amplitude at the $FM$/$I$ interface (i.e. $n = -N_F$ - see 
Fig. \ref{Fig1}). If $\chi_{-N_F}$ changes its sign, the spontaneous current is 
generated. This happens exactly for the same values of the exchange splitting, 
for which a band of the Andreev bound states crosses the Fermi energy. 
Moreover, the presence or absence of the current has a dramatic effect on 
the pairing amplitude $\chi_{-N_F}$. This can be seen in the Fig. \ref{Fig13}, 
where $\chi_{-N_F}$ is plotted as a function of the dimensionless parameter 
$\Theta = 2.79 d E_{ex}/\pi t$.
As implied by Eq.\ref{Delta}, $\chi_n$ is the real part of a complex order
parameter which has an amplitude $\bar{\chi}_n$ and a phase $\bar{\varphi}_n$.
Clearly, Fig. \ref{Fig13} can be red as $\bar{\varphi}_{-N_F}$ flipping from 
$\varphi^0_{-N_F}$ where $\chi_{-N_F}$ has the same sign as $\chi$ in $SC$, to 
$\varphi_{-N_F} = \pi$, where $\chi_{-N_F}$ has the opposite sign.

Evidently, the pairing amplitude is pinned to zero by the spontaneous current 
in the system. Introducing the dimensionless quantity $\Theta$ we can say, that 
system is in a $\pi$-like phase (for odd $\Theta$) or in the $0$-phase 
(even $\Theta$), in close analogy to studies on the $\pi$-effect in the 
$SC$/$FM$/$SC$ structure by Chtchelkatchev et al. \cite{Chtchelkatchev}. 
In Ref. \cite{KGA}, we have shown a similar picture for the thickness of the 
$FM$ slab equal to $10$ layers. Here we show the corresponding results for 
$N_F = 20$ layers. To be precise in the present case $\Theta$ has slightly
different value with prefactor $2.79$ (not $3$ as it was for $10$ layers case).
This discrepancy may come from the fact, that effective exchange splitting is 
little bit smaller than its bulk value (see discussion in the paragraph 
\ref{subsec3.a}).

{\center
\vspace{0.5cm}
\subsection{\label{subsec4.e} Band filling and the polarization of the current}
}

\begin{figure}[h]
 \resizebox{8cm}{5cm}{
  \includegraphics{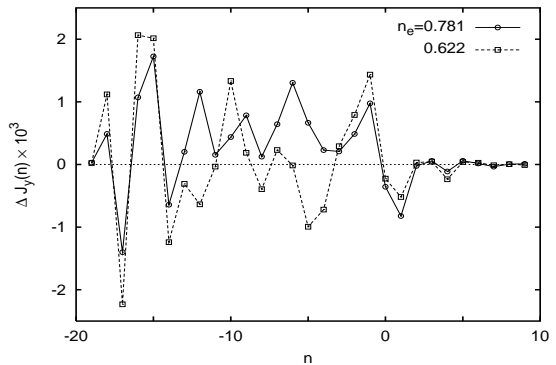}}
 \caption{\label{Fig14} Spin polarized current 
          $\Delta J_y(n) = J_{y\uparrow}(n) - J_{y\downarrow}(n)$ as a 
	  function of the layer index $n$ for exchange splitting $E_{ex}=0.44$
	  and concentration of the electrons $n_e = 0.781$ (solid line) and
	  $0.622$ (dashed line).}
\end{figure}

As our system consists of ferromagnet and superconductor, we can expect the
spontaneous current to be polarized. However, up to now, we have presented 
results for the special case of the particle - hole symmetry ($n_e = 1$ or 
$\mu = 0$) and we found that the spontaneous current was, in fact, not 
spin - polarized. The reason is as follows. Within linear response theory, 
the total current can be divided into two parts: a diamagnetic one - giving the 
response of the bulk density, and a paramagnetic one, which is proportional to 
the density of states at the Fermi energy, as it comes from the deformation 
of the wave function at the Fermi surface \cite{Fauchere}. Now if we note, 
that for the particle - hole symmetric case ($n_e = 1$), the spin up 
$\rho_{tot\uparrow}(\varepsilon_F)$ and spin down 
$\rho_{tot\downarrow}(\varepsilon_F)$ $DOS$ at the Fermi energy are equal, we 
are not surprised that the polarization of the current is equal to zero. As 
we go away from the half filling, the difference between spin up and spin down 
$DOS$ 
($\Delta \rho_{tot}(\varepsilon_F) = \rho_{tot\uparrow}(\varepsilon_F) -
\rho_{tot\downarrow}(\varepsilon_F)$ starts to play a role as it leads to the
polarization of the current. A typical example of the polarization of the 
current $\Delta J_y(n) = J_{y\uparrow}(n) - J_{y\downarrow}(n)$ is shown in the 
Fig. \ref{Fig14} for concentration of electrons $n_e = 0.781$ and $0.622$. 
To make the two calculations comparable we have adjusted the parameters so the 
value of the $SC$ order parameter would be the same. This was achieved by 
shifting of the coupling constant $U_S$ to $-2.129$ for $n_e=0.781$, $-2.345$ 
for $n_e=0.622$ as well as chemical potential to $\mu = - 0.5$ and $- 1.0$ 
respectively.

The total polarization of the current $\Delta J^{tot} = \sum_n J_y(n)$ is of
order $10^{-3}$ and does not depend much on the concentration of electrons
$n_e$. Moreover, we have found a very interesting property of this current,
namely, the sign of the polarization of the current depends on whether 
$n_e < 1$ or $> 1$. This means, that for $n_e = 1 + x$ we get the same results
as for $n_e = 1 - x$, except of the sign of the polarization of the current.
This effect can be explained very easily if we recall that polarization of 
the current is proportional to the difference in the $DOS$ at the Fermi energy, 
and one can check, that $\Delta \rho_{tot}(\varepsilon_F)$ is an 
antisymmetric function under the shift $1-x \rightarrow 1+x$, i.e.
$\Delta \rho_{tot}(\varepsilon_F,1-x) = -\Delta \rho_{tot}(\varepsilon_F,1+x)$.

\vspace{1cm}
%%%%%%%%%%%%%%%%%%%%%%%%%%%%%%%%%%%%%%%%%%%%%%%%%%%%%%%%%%%%%%%%%%%%%%%%%%%%%%

{\center
\section{\label{sec5} HOW TO OBSERVE THE CURRENT AND ITS POLARIZATION ?}
}

Clearly, the important issue raised by the above results is ``how to observe 
experimentally the predicted current ?'' But first of all let's estimate how 
big this current is. If we assume a bandwidth equal to $1 \; eV$, the average 
current per layer is of order of tenths of $\mu A$. In general the 
polarization of the ferromagnet can be large, however it cannot be equal to 
$1$. There must be a small number of the minority electrons. In other words it 
doesn't matter how big the ratio $\xi_F/\xi_S$ is, we should always observe 
the current. The next point is the thickness of the $FM$ region. Namely, the 
current is bigger for a thinner sample, but even for thickness of order of 
several hundreds of $\xi_F$, we expect the current to be easily detectable.

Clearly, the presence of the current should manifest itself in a number of 
experiments. For example, since there is a net magnetic flux associated with 
the current, it could be detected by $SQUID$ experiments. Another possibility 
is to measure the conductance across the $FM$/$SC$ junction \cite{Kashiwaya1}. 
In such experiments the splitting of the zero energy state in the $DOS$ is 
expected to be observed and, as indicated by Eq.\ref{splitting} the energy 
distance between the split peaks measures the current flowing. Finally one can 
also use some local probes like $STM$ techniques to measure the layer resolved 
$DOS$.

Confirming that the current is polarized is more troublesome. 
As we have shown, to get polarized current the system has to be away from the 
$n_e = 1$ point. In our case, to avoid a Fermi energy mismatch, both the $FM$ 
and $SC$ are moved from $n_e = 1$. Fortunately, this is only necessary for the 
ferromagnet. The superconductor can be in a particle - hole symmetric state. In
short, realistic estimations of the spin polarization of the spontaneous 
current require more material specific calculations.

\vspace{1cm}
%%%%%%%%%%%%%%%%%%%%%%%%%%%%%%%%%%%%%%%%%%%%%%%%%%%%%%%%%%%%%%%%%%%%%%%%%%%%%%

{\center
\section{\label{sec6} CONCLUSIONS}
}

In conclusion, we have studied properties of the insulator (or vacuum) - 
ferromagnet - superconductor trilayer. We have shown, that such a structure
supports Andreev bound states forming Andreev bands, whose position can be 
tuned by thickness of the sample or exchange splitting. Moreover, when a band 
crosses the Fermi energy, a spontaneous current (and magnetic flux) is 
generated in the ground state. We have found a relation between the pairing 
amplitude in the ferromagnet, which has oscillatory behaviour, and the Andreev 
bound states. Namely, when the Andreev band crosses the Fermi energy the 
pairing amplitude at the surface opposite to the $FM$/$SC$ interface changes 
sign, and as long as current flows it is pinned to zero. The polarization of 
the current strongly depends on the band filling and is related to the 
difference in the spin up and spin down density of states at the Fermi level. 
We also have discussed the question of the supercurrent vs quasiparticle 
current and gave some experimental clues on how to observe that current. The 
presented results can be, and we have argued that they are, attributed to the 
$FFLO$ effect influenced Andreev bound states and can be verified 
experimentally.

\vspace{0.5cm}
%%%%%%%%%%%%%%%%%%%%%%%%%%%%%%%%%%%%%%%%%%%%%%%%%%%%%%%%%%%%%%%%%%%%%%%%%%%%%%

\begin{acknowledgments}
This work has been supported by Computational Magnetoelectronics Research
Training Network under Contract No. HPRN-CT-2000-00143.
\end{acknowledgments}

%%%%%%%%%%%%%%%%%%%%%%%%%%%%%%%%%%%%%%%%%%%%%%%%%%%%%%%%%%%%%%%%%%%%%%%%%%%%%%

\end{document}